\documentclass[aps,prl,twocolumn,floats,showpacs]{revtex4}
\usepackage{graphicx}
\usepackage{amsmath}
\usepackage{amssymb}
\usepackage{units}
\usepackage[usenames,dvipsnames]{color}

\newcommand{\beq}{\begin{equation}}
\newcommand{\eeq}{\end{equation}}
\newcommand{\bea}{\begin{eqnarray}}
\newcommand{\eea}{\end{eqnarray}}

\begin{document}
\title{Coherent anti-Stokes Raman spectroscopy in the presence of strong resonant signal from background molecules}

\date{\today}
\author{M.~Bitter and V.~ Milner}
\affiliation{Department of  Physics \& Astronomy and The
Laboratory for Advanced Spectroscopy and Imaging Research
(LASIR), The University of British Columbia, Vancouver, Canada \\}

\begin{abstract}
Optical spectroscopy with broadband femtosecond laser pulses often involves simultaneous excitation of multiple molecular species with close resonance frequencies. Interpreting the collective optical response from molecular mixtures typically requires Fourier analysis of the detected time-resolved signal. We propose an alternative method of separating coherent optical responses from two molecular species with neighboring excitation resonances (here, vibrational modes of oxygen and carbon dioxide). We utilize ro-vibrational coupling as a mechanism of suppressing the strong vibrational response from the dominating molecular species (O$_{2}$). Coherent ro-vibrational dynamics lead to long ``silence windows'' of zero signal from oxygen molecules. In these silence windows, the detected signal stems solely from the minority species (CO$_{2}$) enabling background-free detection and characterization of the O$_2$/CO$_2$ mixing ratio. In comparison to a Fourier analysis, our technique does not require femtosecond time resolution or time-delay scanning.
\end{abstract}


\maketitle

\section{Introduction}
Time-resolved coherent anti-Stokes Raman scattering (CARS) with ultrashort laser pulses reveals the evolution of molecular vibrational coherence on a femtosecond time scale \cite{Zewail-book}. CARS spectroscopy is widely used for chemical analysis based on the detected vibrational frequencies which often serve as a unique molecular fingerprint. If the sample of interest is a mixture of molecules with  multiple Raman transitions covered by the broad spectrum of the excitation pulses, the resulting CARS signal will reflect the frequency beating of different Raman modes from the molecular components of the mixture. To examine the chemical composition of the sample, time-resolved detection, typically carried out by scanning the time delay between the vibrational excitation and an ultrashort probe pulse, is followed by a Fourier analysis of the observed CARS signal (for comprehensive reviews on CARS spectroscopy and microscopy, see e.g. \cite{Evans2008,Roy2010,El-Diasty2011}).

In this paper we demonstrate a method of detecting one molecular component (hereafter referred to as ``minority species'') in the presence of another component (the ``background gas'') by means of a simple CARS measurement which does not require high frequency or time resolution and can be implemented as a single-shot detection. The proposed approach relies on the effect of ro-vibrational coupling in Raman excitation of the background species. It is based on the earlier observation \cite{Beaud2001,Bitter2012} that the coupling between the vibrational and rotational degrees of freedom results in the appearance of relatively long (tens of picoseconds) ``silence windows'' in the molecular CARS response. During the silence period, background molecules in the ensemble are vibrating out-of-phase with one another which results in a complete suppression of the \textit{resonant} background signal. Similarly to the suppression of non-resonant background, often required in femtosecond CARS, suppressing the dominant resonant background helps to increase the detection sensitivity in the case of gas mixtures.

Ro-vibrational coupling between the coherently excited vibrational states and a thermal bath of rotational states leads to the vibrational dephasing \cite{Hansson2000,Wallentowitz2002, Branderhorst2008,Bartram2010} and a corresponding decay of CARS signal. The latter decay has been used in gas thermometry, since its rate depends on the initial rotational population distribution \cite{Lang2001,Lucht2006,Roy2008,Roy2009,Kulatilaka2011,Yue2012}. On a longer time scale, the dephasing is followed by a series of full and fractional revivals of the ro-vibrational wave packet, analyzed both theoretically  \cite{Hansson2000, Wallentowitz2002} and experimentally \cite{Beaud2001, Bitter2012}. The revivals stem from the discrete energy spectrum of molecular rotation. The duration of a revival has been suggested as an indicator of the gas temperature \cite{Lang1999,Beaud2001, Lang2001}.

In this work, we focus on detecting the mixing ratio of a binary gas mixture rather than its temperature. As in gas thermometry, the decay of vibrational coherence on a short time scale (i.e. first few picoseconds) has been used for detecting molecular concentrations \cite{Lucht2006,Roy2009}. In contrast, we exploit the long-term evolution of the vibrational coherence, which contains time intervals during which the background molecules are ``silent'' whereas the minority molecules are not (i.e. generate zero and non-zero CARS signal, respectively).

Similarly to the recent work by Kulatilaka \textit{et.al.} \cite{Kulatilaka2011} we mix oxygen with carbon dioxide as the two molecular species with close vibrational frequencies. We follow the ro-vibrational dynamics of the background gas (in our case, O$_2$) for hundreds of picoseconds and position our probe pulses in its silence window. We show that within this window, the detected CARS signal is generated solely by the minority species (here, CO$_2$). Without the strong interfering resonant response from oxygen, the presence and approximate concentration of CO$_2$ can be determined without the traditional Fourier analysis of the time-resolved signal, therefore enabling a single-shot detection. Because the required time resolution of the proposed technique is determined by the width of the silence period which lasts for more than ten picoseconds, no special care is needed for supplying ultrashort transform-limited probe pulses. From a practical standpoint, the ratio of O$_2$ to CO$_2$ yields information about the completeness of oxidation in flames and combustion environments \cite{Bengtsson1995, Thumann1997,Reichardt2001}.

The experimental system consists of a Ti:Sapphire femtosecond regenerative amplifier (SpitFire Pro, Spectra-Physics) generating Stokes pulses at 800 nm, and two optical parametric amplifiers (TOPAS, Light Conversion and OPA-800C, Spectra Physics) producing pump and probe pulses at 718nm and 688 nm, respectively. Folded BOXCARS geometry \cite{Shirley1980} is used to spatially overlap all three pulses inside a vapor cell filled with O$_{2}$ and CO$_2$ at room temperature and variable density ratio. A spectrometer is used to detect the anti-Stokes signal at 626 nm.

\begin{figure}
\centering
 \includegraphics[width=1.0\columnwidth]{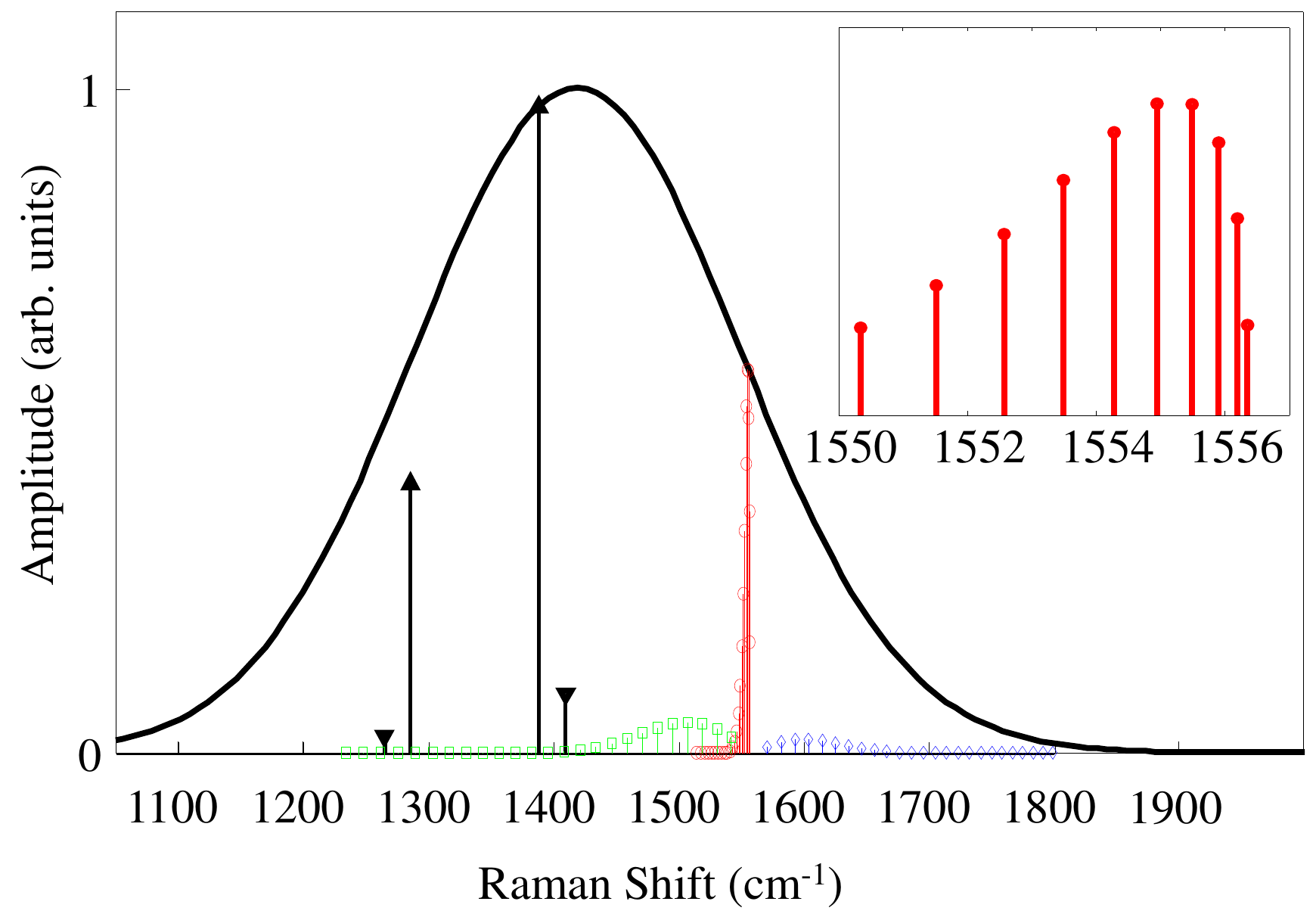}
     \caption{
     (Color online) Calculated two-photon spectrum of the pump-Stokes field (solid black line). $O$, $Q$ and $S$ branches of $v=0 \rightarrow v=1$ transition in O$_2$ are indicated with green rectangles, red circles and blue diamonds, respectively. Using the procedure outlined in our earlier work \cite{Bitter2012}, we found that the relative weight of the $O$ and $S$ branches is about 17 times weaker than that of the $Q$ branch. Key to this work is the rotational splitting of vibrational resonances, shown in the inset for the strongest $Q$ branch. Thermal rotational distribution at room temperature has been taken into account. Two Fermi dyads of CO$_2$ are depicted as black triangles. The relative magnitude of each peak reflects the corresponding relative cross section \cite{Tejeda1995}.}
  \vskip -.1truein
  \label{Fig:Raman-field}
\end{figure}

The two-photon excitation spectrum for the chosen pump and Stokes wavelengths is shown in Fig.~\ref{Fig:Raman-field}. It covers one strong Raman mode of oxygen, corresponding to the $Q$ branch of $v=0 \rightarrow v=1$ vibrational transition centered at 1555 cm$^{-1}$, and one strong Fermi dyad of carbon dioxide at 1285 cm$^{-1}$ and 1388 cm$^{-1}$. The latter is a result of the resonant coupling between the Raman-active symmetric stretch and the first overtone of the bending mode of CO$_{2}$ \cite{Stoicheff1958,Tejeda1995}. Shown in the inset is the rotational splitting of the vibrational resonance in O$_{2}$. For carbon dioxide, this splitting is much smaller \cite{Arakcheev2007a,Arakcheev2007b} and can be neglected for the purpose of this study.

\begin{figure}
\centering
 \includegraphics[width=1.0\columnwidth]{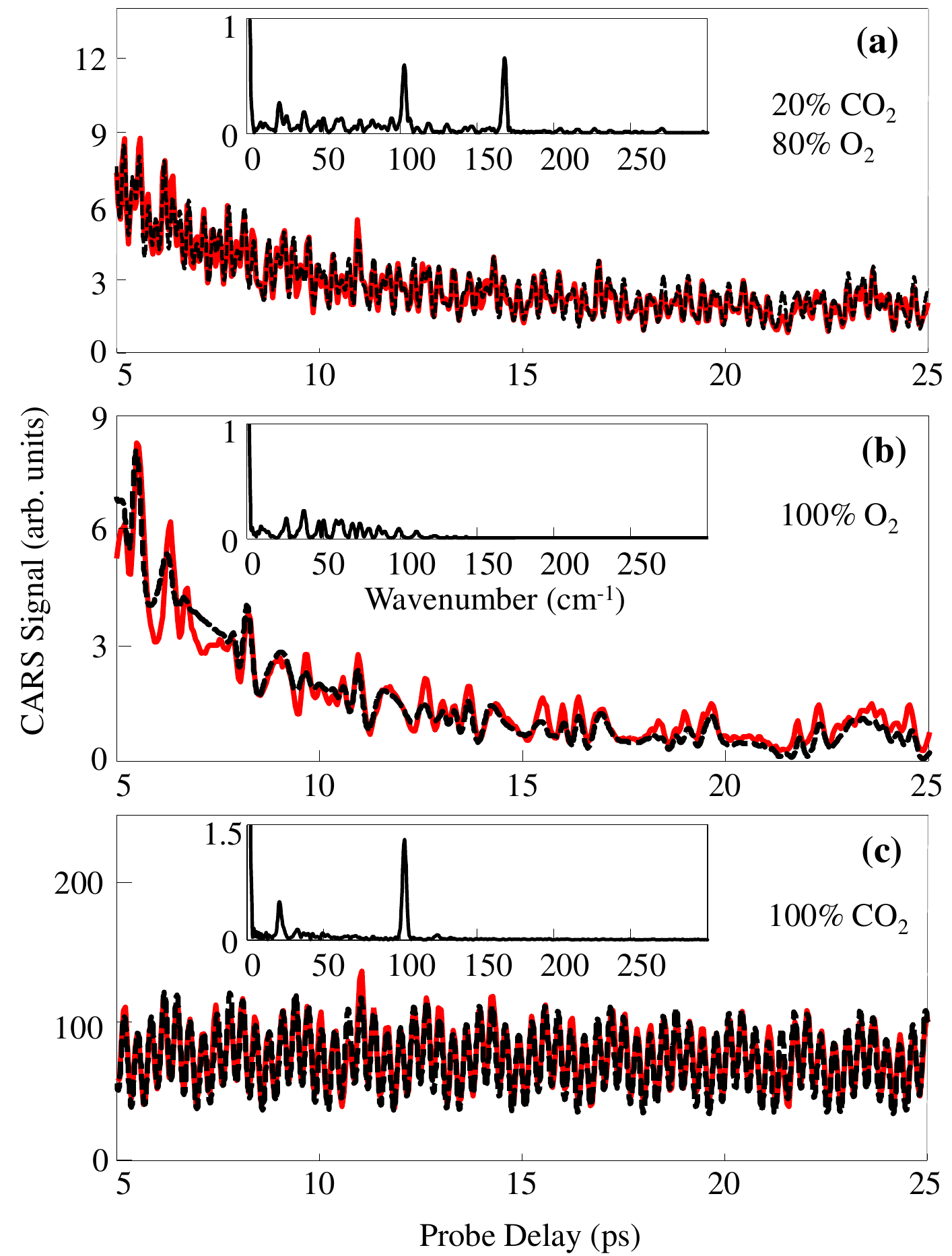}
     \caption{
     (Color online) Time-resolved CARS signal at room temperature from (a) mixture of O$_2$ and CO$_{2}$, (b) pure oxygen, and (c) pure carbon dioxide. Comparing experimental data (solid red line) with numerical calculation (dashed black line) is used for extracting the relative contribution of $O$, $Q$ and $S$ branches in oxygen, as well as the collisional decay time. Respective Fourier spectra are shown in the insets (with different vertical scales).}
  \vskip -.1truein
  \label{Fig:shortscanFFT}
\end{figure}

Figure \ref{Fig:shortscanFFT} shows the observed CARS response from the O$_2$/CO$_2$ mixture (panel a) and its individual components (panels b and c), measured by scanning the time delay between an impulsive 140 fs pump-Stokes excitation and a 140 fs probe pulse. As shown by the Fourier spectra in the insets, the signal of the mixture reflects the 103 cm$^{-1}$ splitting of the strong v$_{1}$/2v$_{2}$ Fermi dyad of carbon dioxide, and the beating at 167 cm$^{-1}$ between the blue dyad line and oxygen vibration (1388 cm$^{-1}$ and 1555 cm$^{-1}$ peaks in Fig.~\ref{Fig:Raman-field}, respectively). The beating between the red dyad line and oxygen vibration (1285 cm$^{-1}$ and 1555 cm$^{-1}$, respectively) at 270 cm$^{-1}$ cannot be observed due to the insufficient temporal resolution. The overall decay on a few-picosecond time scale stems from the rotational dephasing of oxygen vibration (middle panel) which depends on the difference between the rotational constants, $|B_1-B_0| \sim 10^{-2}$ cm$^{-1}$, of the ground and excited vibrational states. The small oscillations are due to the contribution of $O$ and $S$ branches \cite{Bitter2012}.

In contrast to O$_{2}$, the signal from pure CO$_{2}$ is hardly changing on this time scale, as shown in panel c of Fig.~\ref{Fig:shortscanFFT}. The ro-vibrational coupling in carbon dioxide is at least two orders of magnitude weaker, $|B(10^00)-B(00^00)|\sim 10^{-5}$cm$^{-1}$ and $|B(02^00)-B(00^00)|\sim 10^{-4}$cm$^{-1}$ \cite{Arakcheev2007a,Arakcheev2007b}, resulting in a much longer rotational dephasing time of hundreds of picoseconds. Both the decay and the beating pattern in Fig.~\ref{Fig:shortscanFFT}(a) depend on the relative concentration of the two gases in the mixture. Recording long probe scans with high time resolution can therefore provide information about the mixing ratio of interest. Below we present an alternative method, in which we exploit coherent ro-vibrational dynamics on a much longer time scale.

\begin{figure}
\centering
 \includegraphics[width=1.0\columnwidth]{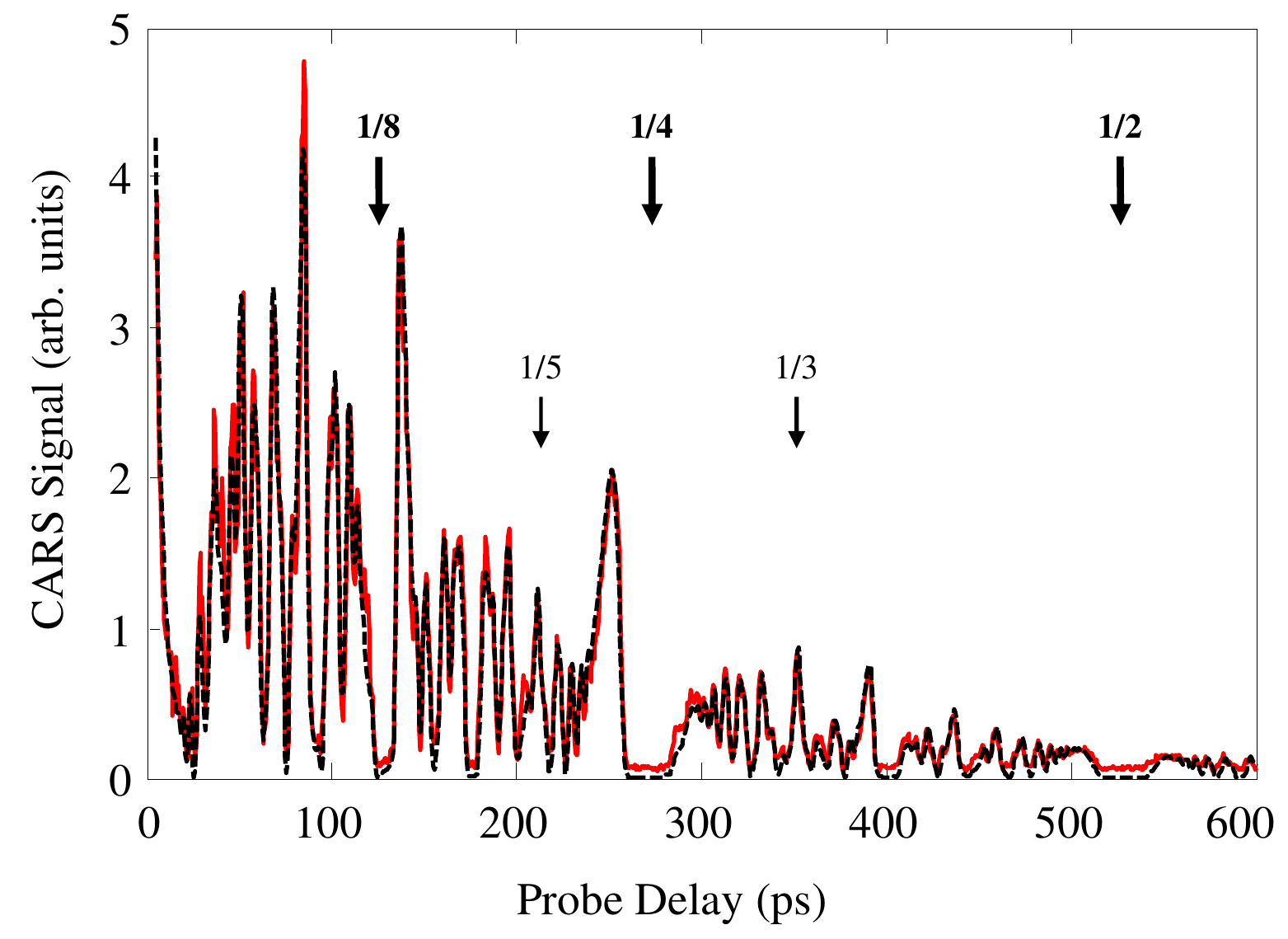}
     \caption{
     (Color online) Dephasing and rephasing of CARS signal from O$_2$ at room temperature and pressure of $300\pm2$ Torr: experimental data (solid red line) and numerical calculation (dashed black line). Vertical arrows point at fractional revivals of the coherently excited ro-vibrational wave packet. The windows of silence can be seen around $1/2,1/4$ and $1/8$ revival time at 530 ps, 265 ps and 130 ps, respectively.}
  \vskip -.1truein
  \label{Fig:Oxygen-longscan}
\end{figure}

Figure \ref{Fig:Oxygen-longscan} shows the long term evolution of the vibrational coherence of pure oxygen up to  600 ps. At a pressure of $300\pm2$ Torr the collisional dephasing time has been determined as $157\pm 3$ ps. The ro-vibrational evolution is governed by the interference between the multiple rotational components of the vibrational transition (inset in Fig.~\ref{Fig:Raman-field}), which results in a sequence of fractional revivals of the ro-vibrational wave packet. The full ro-vibrational revival occurs at $T_\text{rovib} =  1/\left[2c(\alpha_e -2\gamma_e)\right]$, with $c$ being the speed of light in vacuum and $\alpha_e,\gamma_e$ the ro-vibrational coupling constants \cite{NIST}. From our measurements at lower pressure (not shown) we found $T_{rovib}=1.057\pm0.002$~ns. At this moment, all molecules in the ensemble are vibrating in-phase with one another regardless of their angular momentum state. Partial re-phasing of the wave packet is observed at fractional revivals with odd denominator, e.g. $1/5 \times T_{rovib}$ and $1/3 \times T_{rovib}$ marked in Fig.~\ref{Fig:Oxygen-longscan}. Fractional revivals with even denominator, e.g. $1/2,1/4$ and $1/8$ in the figure, correspond to the out-of-phase vibration manifested by the intervals of zero CARS response or ``silence windows''.

\begin{figure}
\centering
 \includegraphics[width=1.0\columnwidth]{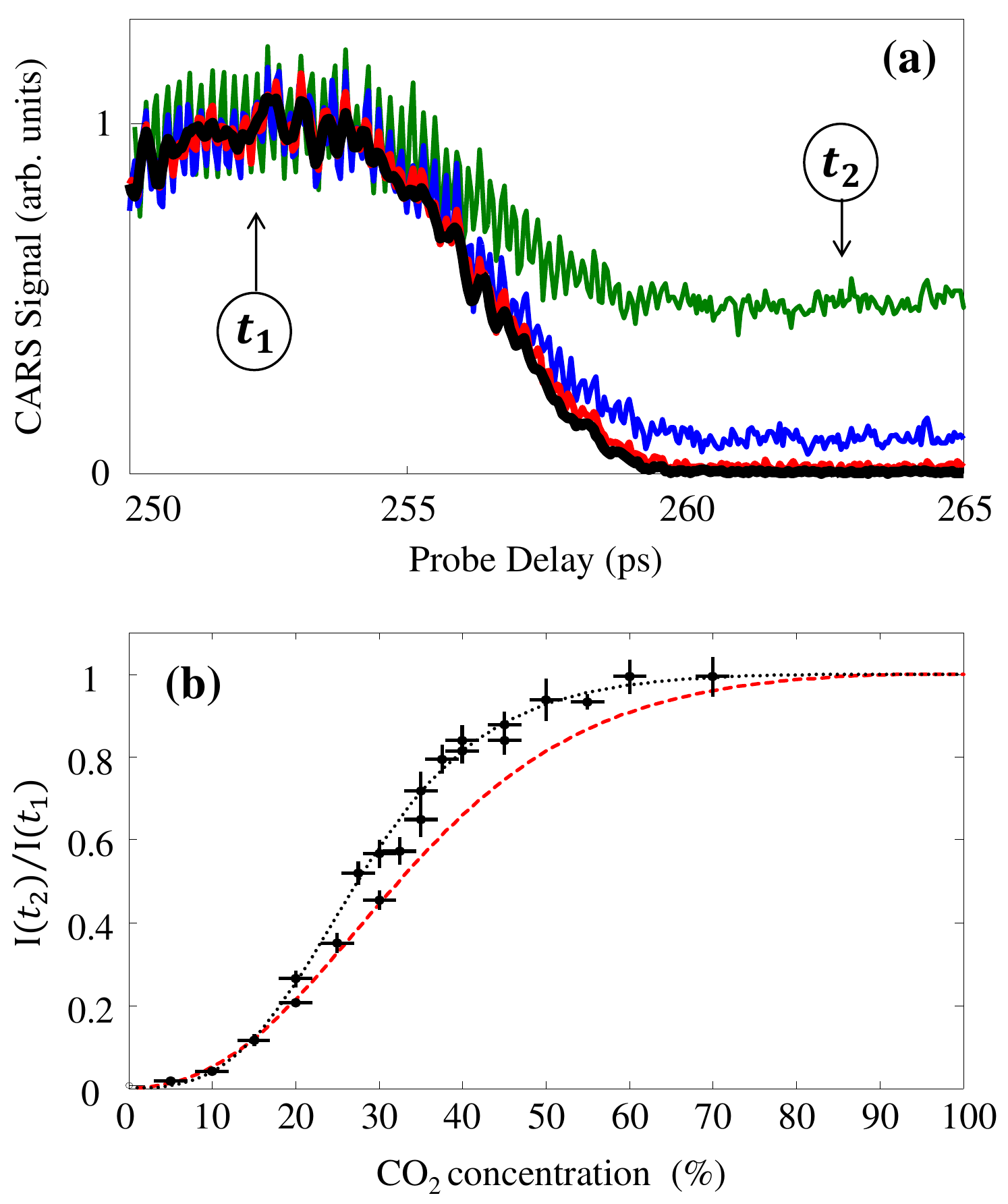}
     \caption{
     (Color online) (a) CARS signal around a silence window of oxygen detected with transform-limited probe pulses of 140 fs duration. Higher curves correspond to higher relative concentration of CO$_{2}$ at 0\%, 10\%, 20\% and 30\% level. All curves have been normalized to their average value around 252 ps. In panel (b), probe pulses have been stretched to 1.3 ps. The intensity ratio of CARS signals at times $t_2$ and $t_1$ is plotted for various CO$_2$ concentrations. The dashed and dotted lines correspond to a simple model and a numerical fit, respectively, explained in the text. }
  \vskip -.1truein
  \label{Fig:Window}
\end{figure}

In silence windows, the strong resonant signal from O$_2$ is largely suppressed, which enables an easy and quick detection of the CO$_{2}$ level. Figure~\ref{Fig:Window}(a) shows a scan around the ro-vibrational quarter-revival time of oxygen at 265 ps. In the case of pure oxygen, the only signal detected after 260 ps is due to the instrumental noise. As the concentration of carbon dioxide is increased, the observed CARS signal is rising. Hence, the mixing ratio can be extracted by carrying out two single-shot measurements, one outside the silence window (e.g. at $t_1=$252 ps) and another one inside it (e.g. at $t_2=$263 ps). The ratio of the signals at times $t_2$ and $t_1$, $I(t_2)/I(t_1)$, can then be used to find the mixing ratio of the molecular concentrations, $R=N_{\mathrm{O}_2}/N_{\mathrm{CO}_2}$, as shown in Fig.\ref{Fig:Window}(b).

Since CARS intensity scales quadratically with the number of molecules \cite{El-Diasty2011}, one might na\"{i}vely expect that with long pulses (i.e. when the interference effects average out) the observed intensity ratio will behave as $1/(1+\alpha R^2)$, with the difference in the transition dipole moments and Franck-Condon factors between CO$_2$ and O$_2$ taken into account by the numerical factor $\alpha $. This model dependence, shown as red dashed line in Fig.\ref{Fig:Window}(b), is in clear disagreement with our experimental data at large CO$_2$ concentrations. We attribute this discrepancy to the fact that the vibrational coherence of different molecular species in a mixture decays differently, with the dephasing rate depending on their collision partners (see \cite{Steinfeld1991} and references therein). As a result, the relative amount of the \textit{vibrationally coherent} molecular components in the mixture does not only depend on the initial mixing ratio $R$ but also changes with time. In the case considered here, i.e. O$_2$/CO$_2$ mixture probed at the quarter-revival silence window of oxygen, we find that the normalized CARS intensity behaves proportionally to $1/(1+\alpha R^{2.7})$ (black dotted line in Fig.\ref{Fig:Window}(b)). An unknown mixing ratio $R$ can therefore be extracted from this empirically determined scaling law.

We note that the proposed detection technique does not require high time resolution as it does not rely on the Fourier analysis of time-resolved data. It can be executed with non-transform-limited probe pulses, as long as their duration is shorter than the width of the silence window. With oxygen as a background gas, this leads to an upper limit of a few picoseconds. The results shown in Fig.\ref{Fig:Window}(b) have been obtained with probe pulses of 1.3 ps duration.

In summary, we propose a new technique for suppressing strong resonant signal from background molecules in coherent anti-Stokes Raman spectroscopy of molecular mixtures. Our method is based on the coupling between the vibrational and rotational degrees of freedom in molecular motion. Ro-vibrational coupling results in the silent intervals, or ``silence windows'', when the molecules are vibrating out-of-phase with one another and interfere destructively in their generation of CARS signal. Hence, the strong resonant response from the dominant background molecules can be effectively suppressed by tuning the timing of probe pulses so as to make them arrive in the sample during the silence window. In this case, the detected CARS signal is generated solely by the rest of the molecular mixture (e.g. by the minority species) and can be easily detected and quantified without the strong interference from the ``silent'' background gas.

We demonstrate a successful implementation of the proposed approach using a mixture of O$_2$ and CO$_{2}$. A small quantity of carbon dioxide is detected in the silence window of oxygen, 265 picoseconds after the impulsive vibrational excitation of the mixture. The sensitivity of the method depends on the collisional de-coherence rate, which determines the signal-to-noise ratio at the time of the silent interval. In our case, this resulted in the ability to detect CO$_{2}$ at relative concentrations higher than $\sim5\%$.

The main advantage of our method lies in its single-shot detection capabilities. Since the resonant response from the primary background gas is eliminated, the detection of the secondary molecular species does not require a Fourier analysis of the time-resolved CARS signal. Neither does it rely on the short duration of probe pulses, as long as they fit within the silence window. We demonstrate this feature by detecting the relative concentration of CO$_{2}$ in a single-shot measurement with probe pulses stretched to 1.3 ps. The concept can be applied to other molecular species with ro-vibrational coupling.

\begin{acknowledgements}
The authors would like to thank E.~A.~Shapiro and M.~Motzkus for valuable discussions.
\end{acknowledgements}



\end{document}